# A quantum ring gyroscope based on coherence de Broglie waves

Byoung S. Ham

School of Electrical Engineering and Computer Science, Gwangju Institute of Science and Technology
123 Chumdangwagi-ro, Buk-gu, Gwangju 61005, South Korea
(Submitted on Nov. 05, 2021; bham@gist.ac.kr)

**In a Mach-Zenhder interferometer (MZI), the highest precision for a measurement error is given by vacuum fluctuations of quantum mechanics, resulting in a shot noise limit[1,2,3,4,5]. Because the intensity measurement in an MZI is correlated with the phase difference, the precision measurement (Δn) is coupled with the phase resolution (Δφ) by the Heisenberg uncertainty principle. Quantum metrology offers a different solution to this precision measurement using nonclassical light such as squeezed light or higher-order entangled-photon pairs, resulting in a smaller Δφ for the same Δn[3,4,5,6,7,8]. Here, we propose another method for the high precision measurement overcoming the Heisenberg photon-phase relation, where the smaller Δφ is achieved by phase quantization in a coupled interferometric system via coherence de Broglie waves (CBWs)[9,10]. For a potential application of the proposed method, a quantum ring gyroscope is presented as a quantum version of the conventional ring laser gyroscope used for inertial navigation and geodesy[11-13].**

Precision measurements are the heart of sensing and metrology[1,2,3,4,5,6,7,8]. In statistics, a standard deviation is proportional to the square root of the number of measurements. The minimum sensitivity of the shot noise in classical physics is caused by the uncertainty principle of quantum physics. This is called a standard quantum limit, which determines the sensing limit in classical physics. The diffraction limit or Rayleigh criterion also classically determines the maximum resolution of sensors. Thus, multi-wave interference in an optical cavity is a typical method to enhance the resolution limit satisfied by coherence optics. In contrast to classical physics, quantum mechanics offers a quantum advantage in sensing and metrology, where higher-order entangled photon pairs play a major role in overcoming the standard quantum limit by a factor of the square root of N, where N is the total number of photons in the entangled pair[3-8]. The higher-order entangled photon pairs is represented by photonic de Broglie waves (PBWs)[14-16]. Due to the indeterminacy and difficulties of PBW generations, however, the implementation of quantum sensing for applications such as lithography[16], frequency standards[17], imaging[18], and spectroscopy[19] has been severely limited.

Quantum mechanics is rooted in the wave-particle duality[20]. Unlike PBWs based on the particle nature of a photon, the wave nature-of coherence de Broglie waves (CBWs) have been recently investigated[21]. The physics of CBWs is in the phase-basis superposition between coupled MZIs[9,10]. Owing to the on-demand control of the geometric scalability of MZIs and the inherent benefit of a single-shot measurement due to the wave nature, CBWs provide new opportunities for quantum sensing to overcome the limitations in both quantum and classical counterparts. Such a quantum feature of CBWs can be applied for various quantum engineering fields of sensing and metrologies. Recently, the first application of CBWs to quantum sensors has been proposed for a CBW Sagnac interferometer, whose resolution is beyond that in conventional cavity interferometers[21]. So far, the Sagnac gyroscope has been implemented for optical[22] and matter-wave[23] interferometry as well as atomic spectroscopy[24] and gravitational wave detection[25]. In particular, the ring laser gyroscope offers a higher sensing capability up to one part of $10^8$ in the earth rotation rate Ω[26,27]. Here, a quantum ring gyroscope based on CBWs is presented, whose sensing capability overcomes the classical physics of the ring laser gyroscope. Compared with CBW Sagnac interferometer[21], the proposed quantum ring gyroscope is immune to environmental phase noises caused by vibrations, temperatures, and air turbulences. Moreover, it can be applied directly to the ring laser gyroscope with a minimal modification.

Figure 1 shows a schematic of the proposed CBW-based quantum ring gyroscope. Figure 1a shows the proposed scheme. Figure 1b is the unfolded scheme of Fig .1a and shows two cavity modes. Figure 1c is an equivalent scheme of Fig. 1a, where the modified region with green MZIs plays the function of phase-basis superposition between counter-propagating fields (see the yellow MZIs). The green dotted region of Fig. 1a with a nonpolarizing 50/50 beam splitter, a path-length controller (piezoelectric transducer 1, PZT 1), and a pair of cavity mirrors represents a modified scheme of a ring laser gyroscope for the CBW applications. Here, PZT 1 represents a control parameter of the phase φ to control the cavity length. PZT 2 is another control parameter of the quantum ring gyroscope, where the phase ζ is for the ring cavity condition with the asymmetric (counter propagating light-caused ±ψ) MZI configuration[9]. The ψ-asymmetric MZI configuration is automatically accomplished by the Sagnac effect for the counter-propagating fields[28,29]. Compared with the original CBW scheme, this ±ψ configuration is an essential part of the proposed scheme.

According to the original CBWs[9], the basic building block is composed of the green-yellow MZIs as denoted by 'p' number[10]. In the yellow MZIs, however, PZT 2 ($\zeta$) caused by environmental noises such as vibrations, temperatures, and air turbulences does not affect the Sagnac effect due to exact phase cancelation by the counter-propagating fields. In the modified region, there is no net Sagnac effect, either. Thus, the Sagnac effect in the proposed scheme is environmental noise immune as in the ring laser gyroscope[11-13]. As a result, any rotation rate $\Omega$ induces a relativistic time delay between the counter-propagating fields inside the ring, resulting in the Sagnac effect as in the original ring laser gyroscope. Meanwhile, the $\Omega$-induced Sagnac effect is neglected in the following analysis of the proposed stand-still quantum ring gyroscope for simplicity to prove the quantum gain in phase resolution.

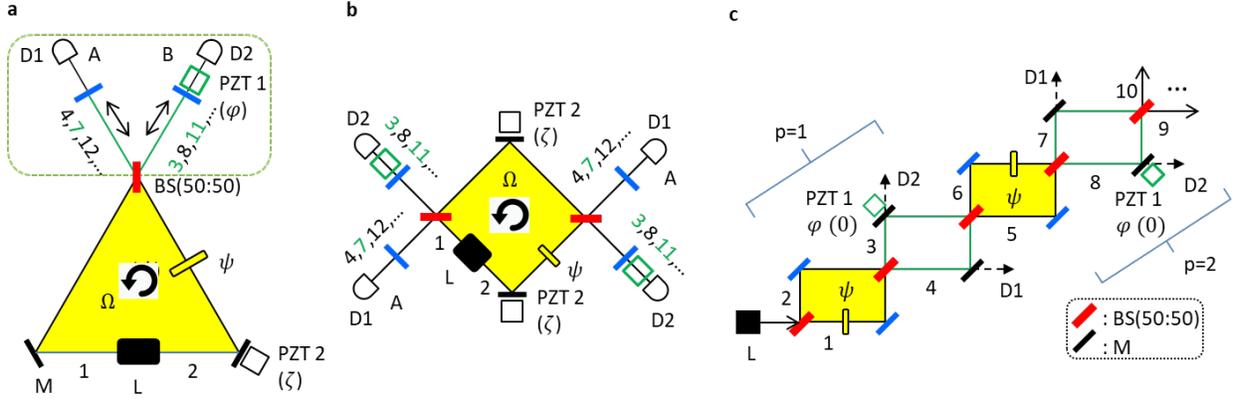

**Figure 1| Schematic of quantum ring gyroscope. a**, Schematic of ring cavity of CBW. **b**, Unfolded scheme of **a**. **c**, Equivalent scheme of CBWs. L: laser, D: photo-detector, BS: 50/50 nonpolarizing beam splitter, M: mirror, PZT: piezo-electric transducer, $\Omega$: rotation rate. The green dotted area with a BS is the modification for CBWs. Both $\varphi$ and $\zeta$ are control parameters for the quantum ring gyroscope. The $\psi$ is the $\Omega$-induced Sagnac effect. Due the inserted BS, each counter-propagating (solid and dotted arrows) field pair has both transmitted (black arrow) and reflected (blue arrows) components. The numbers are sequence of the light propagation across the BS.

Using coherence optics of a BS[30], the output fields A and B in Fig. 1a for the round trip of the cavity are obtained via matrix representations for Fig. 1c as follows:

$$\begin{bmatrix} E_7 \\ E_8 \end{bmatrix} = [MZI]_+[\varphi][MZI]_- \begin{bmatrix} E_0 \\ 0 \end{bmatrix}$$
$$= (-1)^1 e^{i\psi} E_0 \begin{bmatrix} \cos\psi \\ \sin\psi \end{bmatrix}, \quad (1)$$

where $[MZI]_+ = \frac{1}{2}\begin{bmatrix} 1 - e^{i\psi} & i(1 + e^{i\psi}) \\ i(1 + e^{i\psi}) & -(1 - e^{i\psi}) \end{bmatrix}$ and $[MZI]_- = \frac{1}{2}e^{i\psi}\begin{bmatrix} 1 - e^{-i\psi} & i(1 + e^{-i\psi}) \\ i(1 + e^{-i\psi}) & -(1 - e^{-i\psi}) \end{bmatrix}$, respectively. For simplicity, a laser gain in the ring cavity has not been considered, where $E_0$ is determined by the embedded laser L. The phase $\psi$ is due to the rotation ($\Omega$)-induced Sagnac effect, and $[\varphi] = \begin{bmatrix} 1 & 0 \\ 0 & 1 \end{bmatrix}$ is set with $\varphi = 2n\pi$ (n=1,23,...) by controlling PZT 1. The PZT 2 is for the ring cavity, where the phase $\zeta$ is invariant to the Sagnac effect. Unlike the original CBWs, the basic building block in Fig. 1c is the $\psi$-$\varphi$ doubly coupled MZIs due to the round trip in the cavity. The global phase of $e^{i\psi}$ in equation (1) has no effect on measurements. For CBWs, however, $\varphi = 0$ must be satisfied, otherwise $E_7 = -e^{i\psi}E_0$ and $E_8 = 0$. Thus, the general solution of the m$^{th}$ order CBW is as follows (m=2p):

$$\begin{bmatrix} E_A \\ E_B \end{bmatrix}^m = [CBW]^m \begin{bmatrix} E_0 \\ 0 \end{bmatrix}$$
$$= (-1)^m e^{im\psi} E_0 \begin{bmatrix} \cos(m\varphi) \\ \sin(m\varphi) \end{bmatrix}, \quad (2)$$

where $[CBW]^m = (-1)^m e^{im\psi} \begin{bmatrix} \cos\psi & -\sin\psi \\ \sin\psi & \cos\psi \end{bmatrix}^m$, resulting in $E_A^{(m)} = (-1)^m e^{im\psi} \cos(m\psi) E_0$ and $E_B^{(m)} = (-1)^m e^{im\psi} \sin(m\psi) E_0$. This is the phase-basis quantization of CBWs, $\psi_m \in \{0, \frac{\pm\pi}{m}\}$[10].

Regarding the output intensities $I_A$ and $I_B$ in Fig. 1a detected by D1 and D1, respectively, result from via linear superposition of all modes, where the corresponding amplitudes are $E_A = E_0 \sum_m (-1)^m \sin(m\psi)$ and $E_B = E_0 \sum_m (-1)^m \cos(m\varphi)$. For this, the following analytical solutions are obtained.

(i)   For $m\psi = \pm\frac{(2n-1)\pi}{2}$, where n=1,2,3…

For all m, $E_B^{(m)} = 0$. $E_B^{(m)} = 0$ is also satisfied due to the $(-1)^m$ effect. Thus, $I_A = I_B = 0$.

(ii)   For $m\psi = \pm 2n\pi$, where n=1,2,3…

For all m, $E_A^{(m)} = 0$ at $\psi = \pm 2n\pi$ due to the $(-1)^m$ effect, resulting in destructive interference. $E_B^{(m)} = 0$ is also satisfied. Thus, $I_A = I_B = 0$.

(iii)   For $m\psi = \pm(2n-1)\pi$, where n=1,2,3…

For all m, all $E_B^{(m)}$ constructively interferers at $\psi = \pm(2n-1)\pi$ due to the $(-1)^m$ effect. However, all $E_B^{(m)} = 0$. Due to the constructive interference, $I_A = \eta I_0$, where the η is the cavity gain (not included). Thus, we need to clarify whether equation (2) is rooted in coherence optics of multi-wave interference or quantum optics of the phase-basis quantization to show the novelty of the proposed quantum ring gyroscope.

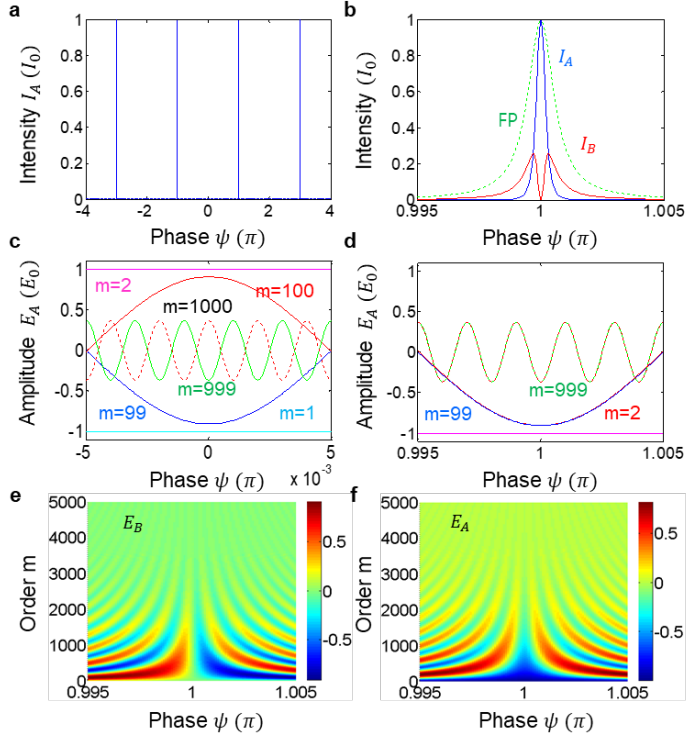

**Figure 2| Numerical calculations for equation (2). a** and **b**, Output intensities. **c – f**, Output amplitudes. Amplitudes and intensities are normalized. Reflection coefficient is set at r=0.999. The ring laser gain is not included. FP: Fabry-Perot interferometer.

For the detailed analysis, numerical calculations are conducted in Fig. 2 using equation (2). Figure 2a shows the normalized output intensity ($I_A$) detected by D1 in Fig. 1. As analyzed above in (i)-(iii), the constructive interference appears at $\psi = \pm(2n-1)\pi$ for $E_A$. This π-shifted position with respect to the conventional cavity optics is due to the inserted BS as shown in Fig. 1b, resulting in a π phase shift between two identical cavities. Figure 2b is an expanded version of Fig. 2a for both output intensities, where $I_B$ is denoted in red. For comparison purposes, the green dotted curve which is π phas shifted shows the classical limit of a Fabry-Perot (FP) interferometer, whose phase resolution is three times worse. Thus, it is clarified that there is a breakthrough in phase resolution in the proposed scheme of quantum ring gyroscope. This breakthrough in Fig. 2b is due to the quantum gain of CBWs.

Figure 2c and 2d illustrate for the destructive and constructive interference in $E_A$. For this, some neighboring samples are shown for the $m^{th}$ and $(m+1)^{th}$ ordered amplitudes in $E_A^{(m)}$. From the symmetric distribution, the destructive interference at $\psi = \pm 2n\pi$ in Fig. 2c is due to the $(-1)^m$ effect in equation (2). On the contrary, there is a constructive interference at $\psi = \pm(2n-1)\pi$ in Fig. 2d as analyzed above. Figure 2e and 2f show all ordered amplitudes up to m=5,000 as a function of $\psi$. As the order m increases, the amplitudes of both $E_A^{(m)}$ and $E_B^{(m)}$ decrease. The sum of amplitudes for all modes of $E_A^{(m)}$ constructively interfere only at $\psi = \pm(2n-1)\pi$. Due to the BS physics, $I_A$ includes the Ω-induced frequency beating.

In conclusion, a quantum ring gyroscope is presented for quantum sensing of CBWs in a ring-type optical resonator. The quantized CBW modes are due to linear superposition of asymmetric MZIs, where the phase quantization corresponds to the energy quantization in PBWs in conventional quantum sensing applications. Compared with cavity optics for a ring laser gyroscope, the proposed quantum ring gyroscope showed a quantum gain in the phase resolution, where this breakthrough cannot be obtained by classical means. Although the physics belongs to many-wave interference in coherence optics, the many waves of CBWs are for the quantized wavelengths. The estimated phase resolution is at least three times greater than that of the conventional cavity-optics interferometer. The present method can also be applied for gravitational wave detection by replacing the squeezed light for better sensitivity (discussed elsewhere).

**Methods**

Figure 2 is calculated by a MATLAB software for equation (2). The MATLAB code is available upon reasonable request to the corresponding author.

**Acknowledgement**

This work was supported by the GIST-GTI 2021 and the ICT R&D program of MSIT/IITP (2021-0-01810), Development of Elemental Technologies for Ultra-secure Quantum Internet.

**Author contributions:** B.S.H. solely wrote the manuscript text with analytic and numerical solutions. Correspondence and request of materials should be addressed to BSH (email: bham@gist.ac.kr).

**Funding:** GIST GRI 2021 and ICT R&D program of MSIT/IITP (2021-0-01810)

**Competing interests:** The author declares no competing interests.